\documentclass[]{jpsj3}
\usepackage{graphicx}
\usepackage{dcolumn}
\usepackage{bm}

\def\Vec#1{\bm{#1}}
\newcommand{\cH}{\mathcal{H}}
\newcommand{\imu}{{\rm i}}
\newcommand{\rme}{{\rm e}}
\newcommand{\rmd}{{\rm d}}

\title{Efficient Numerical Self-Consistent Mean-Field Approach for Fermionic Many-Body
Systems by Polynomial Expansion on Spectral Density}

\author{Yuki \surname{Nagai}, Yukihiro \surname{Ota}\thanks{Present address: Advanced Research Institute, RIKEN, 2-1 Hirosawa, Wako-shi, Saitama 351-0198, Japan}, and Masahiko \surname{Machida} 
}
\inst{
\address{CCSE, Japan  Atomic Energy Agency, 5-1-5 Kashiwanoha, Kashiwa, Chiba 277-8587, Japan} \\
\address{CREST(JST), 4-1-8 Honcho, Kawaguchi, Saitama, 332-0012, Japan}
}
\abst{
We propose an efficient numerical algorithm to solve 
Bogoliubov de Gennes equations self-consistently for inhomogeneous superconducting systems with a reformulated polynomial expansion scheme.
This proposed method is applied to typical issues such 
 as a vortex under randomly distributed impurities and a normal conducting junction sandwiched between
superconductors.
With various technical remarks, we show that its efficiency becomes remarkable in large-scale parallel performance.
}

\kword{mean-field theory, superconductivity, Bogoliubov-de Gennes equations, parallel computation}

\begin{document}
\maketitle

\section{Introduction}

The mean-field approach is one of the most convenient and efficient ways to 
clarify relationship between microscopic descriptions and macroscopic 
phenomena in condensed matter physics. 
So far, this approach has been applied to various
fermionic interacting many-body systems.
Recently, inhomogeneous superconducting systems such as nano-scale superconducting wires and 
artificial/intrinsic Josephson junctions have been its intriguing application examples\cite{Bonca,Inhomo,Tinkham:1996,Ota}.
The Bogoliubov-de Gennes (BdG) equation\cite{deGennes:1999} is a 
theoretical starting-point in these systems.
Its self-consistent solution gives information such as non-trivial quasi-particle excitation-spectra and  inhomogeneous superconducting gap. 
However, such a mean-field BdG approach has been regarded to 
be not practical for a long time, since its calculations require huge computation resources beyond 
the contemporary standard level of present computers.

Very recently, a few groups \cite{Covaci,Zha;Peeters:2010,Han} have proposed a highly
efficient numerical method to solve the BdG equations by using the kernel polynomial expansion. \cite{Weisse} 
In these papers, the key idea is to expand Green's function with a set of the Chebyshev polynomials.
The polynomial expansion drastically reduces the computation cost and has an excellent parallel efficiency 
in contrast to the conventional way, i.e., direct full diagonalization of the BdG-Hamiltonian. 
In the history of condensed matter physics, 
such a Chebyshev-based expansion of Green's function goes back to
seminal studies by Tanaka {\it et al.}\,\cite{Kunishima,Tanaka}. 
These authors calculated Green's functions for a generic Hamiltonian 
with use of various kinds of orthonormal polynomials. 
Sota {\it et al.}\,\cite{Sota} developed this idea as an
oscillation-free Fourier expansion scheme. 
Afterwards, several papers have been published 
in terms of the polynomials expansion of Green's function\cite{Weisse,Okumura}. 
Thus, the polynomial-based expansion has been
an attractive numerical method in large-scale
fermionic many-body systems. 
Specifically, the application to inhomogeneous
superconducting systems is an important target
since spatial profile of the
superconducting gap has key information to predict
their physical properties.

The aim of this paper is to develop a fast and tractable method to
self-consistently calculate the BdG equations for various
inhomogeneous superconducting systems. 
We reformulate the polynomial expansion scheme in a more comprehensible manner 
and perform first self-consistent calculations on typical inhomogeneous superconducting systems. 
At first, we claim that the polynomial expansion is applied 
to not  Green's function itself but the spectral density of Green's function.
By expanding the Dirac's delta function on the spectral density,  
we obtain a full mean-field calculation scheme. 
The present tool, {\it spectral-density polynomial expansion} allows a
straightforward numerical calculation with the mean-fields, i.e., 
superconducting gap and/or general Hartree-Fock terms. 
Next, we claim that the mean-field itself converges faster than the local density
of states.
We focus on $s$-wave superconductivity
to demonstrate its efficiency. 
We remark that our approach can be effective for more general cases including
$d$-wave superconductivity in magnetic fields\cite{NagaiQPI}
and multi-orbital superconductivity with spin-orbit
coupling.

This paper is organized as follows. 
The polynomial-expansion reformula is given in Sec.~II. 
The present style is theoretically more complete and easier to do programing on parallel cluster machines. 
In Sec.~III., we demonstrate three typical examples; an $s$-wave disordered superconductor in 
the magnetic field, i.e., a vortex formation under randomly distributed impurities, 
a nano-scale superconductor-normal-superconductor (SNS) junction, i.e., 
superconducting proximity effects, and  
a numerical challenge, i.e., a reproduction of full temperature dependence 
of the gap amplitude from zero to $T_c$. 
Some convenient technical remarks for practical simulations 
are given in Sec.~IV.  
Section V is devoted to the summary.

\section{Formulation}

We give a full formula based on the orthonormal-polynomial expansion to 
solve the BdG equations.
The essence is that the spectral density of Green's functions
can be expanded by orthonormal polynomials. 
Afterwards, we show the explicit expressions with use of the Chebyshev polynomials. 
The present formalism can be applied 
to any fermionic quadratic Hamiltonian with mean-fields.

\subsection{Hamiltonian}

We start with the Hamiltonian associated with the BdG equations. 
Covaci\,{\it et al.}\cite{Covaci} proposed a way to calculate the eigenvalues and the eigenvectors of this Hamiltonian without full diagonalization. 
We describe their formula in a more general way.

Let us consider a Hamiltonian for a fermion system given as 
\(
H = \Psi^{\dagger}\hat{\mathcal{H}}\Psi/2
\). 
The column vector $\Psi$ is composed of $N$ fermionic annihilation
$c_{i}$ and creation operators $c_{i}^{\dagger}$ ($i=1,\,2,\ldots,\,N$),  
\(
\Psi=(\{c_{i}\},\{c_{i}^{\dagger}\})^{\rm T}
\), where 
\(
\{c_{i}\}
=(c_{1},\,c_{2},\ldots,\,c_{N})^{\rm T}
\) and  
\(
\{c_{i}^{\dagger}\}
=(c_{1}^{\dagger},\,c_{2}^{\dagger},\ldots,\,c_{N}^{\dagger})^{\rm T}
\).
The row vector $\Psi^{\dagger}$ is also defined as 
\(
\Psi^{\dagger}= (\{c_{i}^{\dagger}\}^{\rm T},\{c_{i}\}^{\rm T})
\). 
The symbol $\mbox{T}$ means transposition. 
The fermionic canonical anti-commutation relation leads \(
[c_{i},c_{j}^{\dagger}]_{+}=\delta_{ij}
\). 
The subscription $i$ in $c_{i}$ or $c_{i}^{\dagger}$ indicates a quantum 
index depending on spatial site, spin, orbital, etc. 
The ``Hamiltonian'' matrix $\hat{\mathcal{H}}$ is a $2N\times 2N$ Hermite
matrix given as
\begin{equation}
 \hat{\mathcal{H}}
=
\left(
\begin{array}{cc}
\hat{A} & \hat{B}\\
\hat{B}^{\dagger} & -\hat{A}^{\rm T}
\end{array}
\right),
\end{equation}
where $\hat{A}$ and $\hat{B}$ are $N\times N$ complex matrices.  
These matrices have the relation given as 
\begin{equation}
\hat{A}^{\dagger}=\hat{A},
\quad
\hat{B}^{\rm T} = -\hat{B},  
\end{equation}
because of the hermitian property of $H$ and the fermionic canonical
anti-commutation relations.
When we consider a superconductor, $\hat{\mathcal{H}}$ corresponds to the
mean-field Bardeen-Cooper-Schrieffer (BCS) Hamiltonian and $\hat{B}$ 
contains the superconducting gap. 

\subsection{BdG equations}
The BdG equations are regarded as th eigenvalue equation with respect to
$\hat{\mathcal{H}}$ expressed as 
\begin{subequations}
\begin{eqnarray}
&&
 \hat{\mathcal{H}}\Vec{f}_{(\gamma)} = \epsilon_{\gamma} \Vec{f}_{(\gamma)}
\quad
(\gamma=1,2,\ldots, 2N), \\
&&
 \Vec{f}_{(\gamma)}
=\left(
\begin{array}{c}
\Vec{u}_{(\gamma)} \\
\Vec{v}_{(\gamma)}
\end{array}
\right).
\end{eqnarray}
\end{subequations} 
The column vectors $\Vec{u}_{(\gamma)}$ and $\Vec{v}_{(\gamma)}$ are
$N$-component complex vectors. 
To solve the BdG equations is equivalent to 
diagonalization of $\hat{\mathcal{H}}$ with a unitary
matrix $\hat{U}$ (Ref.~\citen{vanHemmen:1980}),
\begin{align}
\hat{U}^{\dagger} \hat{\cH} \hat{U} &= \hat{D},
\quad
 \hat{D} = {\rm diag}(\epsilon_{1},\epsilon_{2},\ldots,\epsilon_{2N}).
\end{align}
The eigenvalues $\epsilon_{c}$'s are not independent of each other, i.e.,  
\(
 \epsilon_{i} = -\epsilon_{i+N}
\) ($i=1,\,2,\ldots,N$).  
The matrix elements of $\hat{U}$ lead as 
\begin{align}
U_{i \gamma} = u_{(\gamma),i}
\quad
U_{i+N \gamma} = v_{(\gamma),i}. 
\end{align}

\subsection{Spectral density}
Here, we concentrate on a key quantity, i.e., spectral density (or discontinuity)
$\hat{d}(\omega)$, which is a $2N\times 2N$ matrix, to solve the BdG
equations using orthonormal polynomials. 
All essential physical observables are described by bilinear forms with respect to
$\hat{d}(\omega)$.

Now, let us define the Green's function as 
\(
\hat{G}(z) = (z - \hat{\mathcal{H}})^{-1}, 
\)
which is a $2N\times 2N$ complex matrix. 
With the use of the unitary matrix $\hat{U}$, each component of $\hat{G}(z)$ is expressed 
as  
\begin{align}
G_{\alpha \beta}(z) &= \sum_{\gamma = 1}^{2N} U_{\alpha \gamma} U_{\beta \gamma}^{\ast} \frac{1}{ z -
 \epsilon_{\gamma}}
\quad
(1\le \alpha,\beta \le 2N). 
\end{align}
If we set $z = \imu \omega_{n}$ with the Matsubara frequency $\omega_{n} = (2 n +1)/\beta$, 
the above formula corresponds to the temperature Green's function\cite{Mahan:2000}.  
The retarded and the advanced Green's functions are, respectively, defined as
\begin{subequations}
\begin{align}
 \hat{G}^{\rm R}(\omega) &= \lim_{\eta \to 0+}\hat{G}(\omega + \imu \eta), \\
 \hat{G}^{\rm A}(\omega) &= \lim_{\eta \to 0+}\hat{G}(\omega - \imu \eta). 
\end{align}
\end{subequations} 
The spectral density\cite{Mahan:2000} is given as a difference between
the retarded and the advanced Green's functions, 
\(
\hat{d}(\omega) \equiv \hat{G}^{\rm R}(\omega) - \hat{G}^{\rm A}(\omega)
\), whose matrix elements are expressed as   
\begin{align}
\left[ \hat{d}(\omega) \right]_{\alpha \beta} 
&= - 2 \pi \imu \sum_{\gamma = 1}^{2 N} U_{\alpha \gamma} U^{\ast}_{\beta \gamma} 
\delta(\omega - \epsilon_{\gamma}). \label{eq:discon} 
\end{align}
In order to obtain physical observables (e.g., density of states) from
$\hat{d}(\omega)$, we introduce the following useful $2N$-component unit-vectors $\Vec{e}(i)$ and $\Vec{h}(i)$
($1\le i\le N$), which are,
respectively, defined as  
\begin{equation}
 [\Vec{e}(i)]_{\gamma} = \delta_{i,\gamma}, \quad 
 [\Vec{h}(i)]_{\gamma} = \delta_{i+N,\gamma}. 
\end{equation} 
Specifically, employing $\Vec{e}(i)$ and $\Vec{h}(i)$, 
the elements of the column vectors $\Vec{u}_{\gamma}$ and $\Vec{v}_{\gamma}$ are rewritten as 
\begin{align}
u_{(\gamma),i} &= [\Vec{e}(i)^{\rm T} \hat{U}]_{\gamma}, \\
v_{(\gamma),i}^{\ast} &= [\hat{U}^{\dagger} \Vec{h}(i)]_{\gamma}.
\end{align}
%
Furthermore, the local density of states with respect to the site $i$ is given as 
\begin{align}
N(\omega,i) &=- \frac{1}{2 \pi \imu} \Vec{e}(i)^{\rm T} \hat{d}(\omega)
 \Vec{e}(i), \label{eq:ldos} 
\end{align}
with use of $\hat{d}(\omega)$ and $\Vec{e}(i)$, 
since $N(\omega,i)$ is defined as 
\begin{align}
N(\omega,i) 
&= 
\sum_{\gamma = 1}^{2 N} |u_{(\gamma),i}|^{2} \delta(\omega - \epsilon_{\gamma}), \\
&=
\sum_{j=1}^{N}|u_{(j),i}|^{2} \delta(\omega - \epsilon_{j})
+
\sum_{j=1}^{N}|v_{(j),i}|^{2} \delta(\omega + \epsilon_{j}).
\end{align}
%
A typical self-consistent BdG calculation for a superconductor 
requires two types of mean-fields 
\(
\langle c^{\dagger}_{i} c_{j} \rangle
\) 
and 
\(
\langle c_{i} c_{j} \rangle 
\).  
These mean-fields can be expressed as, 
\begin{subequations}
\begin{align}
\langle c^{\dagger}_{i} c_{j} \rangle &= 
- \frac{1}{2 \pi \imu} \int_{- \infty}^{\infty} \rmd \omega f(\omega) 
\Vec{e}(j)^{\rm T} \hat{d}(\omega) \Vec{e}(i), \label{eq:cdc}\\
\langle c_{i} c_{j} \rangle &= 
- \frac{1}{2 \pi \imu} \int_{- \infty}^{\infty} \rmd \omega f(\omega) 
\Vec{e}(j)^{\rm T} \hat{d}(\omega) \Vec{h}(i), \label{eq:cc}
\end{align}
\end{subequations} 
with $f(x) = 1/(\rme^{\beta x} + 1)$. 
Here, $\beta$ is the inverse temperature $\beta = 1/T$.

\subsection{Orthonormal polynomial expansion}
In this paper, we focus on an orthonormal polynomial $\phi_{n}(x)$ with interval [-1,1] ($n=0,1,\ldots$). 
In principle, various orthonormal polynomials are applicable
to solve the BdG equations (as shown in Ref.\,\citen{Tanaka}).
A set of polynomials $\phi_{n}(x)$, which is assumed to be a real function
with respect to $x$, fulfills the relations 
\begin{subequations}
\begin{align}
 \delta(x-x^{\prime}) 
&= \sum_{n=0}^{\infty}\frac{W(x)}{w_{n}}
\phi_{n}(x)\phi_{n}(x^{\prime}), 
\label{eq:delta}
\\
w_{n}\delta_{n,m} &= 
\int_{-1}^{1} \phi_{n}(x)\phi_{m}(x) W(x)\rmd x. \label{eq:wn}
\end{align}
A recurrence formula is generally given as\cite{Abramowitz;Stegun:1972}
\begin{align}
\phi_{n+1}(x) = (a_{n}+b_{n}x)\phi_{n}(x)  -
 c_{n}\phi_{n-1}(x). \label{eq:rec} 
\end{align}
\end{subequations} 
In order to confine the eigenvalue range
inside the interval, we rescale the energy scale of $\hat{\mathcal{H}}$
by the following manner
\begin{align}
 \hat{\mathcal{K}} =  \frac{\hat{\cH} - b \hat{I}}{a},
 \quad
 \xi_{\gamma} =\frac{\epsilon_{\gamma} - b}{a},
\label{eq:rescale}
\end{align}
where 
\(
a =(E_{\rm max}-E_{\rm min})/2
\) and 
\(
b =(E_{\rm max}+E_{\rm min})/2
\) 
with $E_{\rm min} \le \epsilon_{\gamma} \le E_{\rm max}$. 
It seems that these relations require a tough computing task  to
obtain $E_{\rm max}$ and $E_{\rm min}$.  
However, their rough estimations 
are practically enough.
We employ a convenient criterion to determine $E_{\rm max}$ and 
$E_{\rm min}$ from physical information. See Sec.~\ref{subsec:choise_emin_emax} for detail.
 
Now, let us derive a formula using the polynomial expansion.
At first, we define a matrix form by polynomial functions as
\begin{equation}
\left[ \phi_{n}(\hat{\mathcal{K}}) \right]_{\alpha \beta} 
= \sum_{\gamma=1}^{2 N} U_{\alpha \gamma} U_{\beta \gamma}^{\ast}\phi_{n}(\xi_{\gamma}), 
\end{equation}
where $\phi_{n}(\xi_{\gamma})$ is well-defined in the interval  
\(
\xi_{\gamma} \in [-1,1]
\). 
This implies that $\omega$-integrals in Eqs.~(\ref{eq:cdc}) and
(\ref{eq:cc}) are also bound in the finite energy range. 
According to a similar manner to Eq.~(\ref{eq:rescale}), 
these integral intervals become $[-1,1]$, i.e. 
\(
\omega = a x + b
\) with $x\in [-1,1]$. 
Substituting the right hand side of Eq.~(\ref{eq:delta}) for the
definition of $\hat{d}(\omega)$, we have
\begin{align}
\Vec{p}^{\rm T} \hat{d}(\omega) \Vec{q} 
&= 
-\frac{2\pi \imu}{a}
\sum_{n=0}^{\infty}\frac{W(\omega)}{w_{n}}\phi_{n}(\omega)
\,\! \Vec{p}^{\rm T}\Vec{q}_{n},  
\label{eq:blform_d}
\end{align}
for arbitrary $2N$-component real vectors $\Vec{p}$ and $\Vec{q}$.  
A sequence of the vector $\Vec{q}_{n}(\equiv \phi_{n}(\hat{\mathcal{K}})\Vec{q})$
is recursively generated by 
\begin{subequations}
\begin{align}
 \Vec{q}_{n+1} 
&= (a_{n} + b_{n}\hat{\mathcal{K}}) \Vec{q}_{n} -
  c_{n}\Vec{q}_{n-1}
\quad
(n\ge 2), \label{eq:rec_matrix} \\
 \Vec{q}_{1} 
&= \phi_{1}(\hat{\mathcal{K}})\Vec{q}, 
\quad \Vec{q}_{0} 
= \phi_{0}(\hat{\mathcal{K}})\Vec{q}.
\end{align}
\end{subequations} 
The coefficients of Eq.~(\ref{eq:rec_matrix}) are the same as the ones
in Eq.~(\ref{eq:rec}). 
Accordingly, the mean-fields (\ref{eq:cdc}) and (\ref{eq:cc})
are expressed as 
\begin{subequations} 
\begin{align}
\langle c^{\dagger}_{i} c_{j} \rangle &= 
\sum_{n = 0}^{\infty}
\Vec{e}(j)^{\rm T} \Vec{e}_{n}(i) \frac{{\cal T}_{n}}{w_{n}}, \label{eq:cdct}\\
\langle c_{i} c_{j} \rangle &= 
\sum_{n = 0}^{\infty}
\Vec{e}(j)^{\rm T} \Vec{h}_{n}(i) \frac{{\cal T}_{n}}{w_{n}}, \label{eq:cct} 
\end{align}
\end{subequations} 
where
\begin{align}
{\cal T}_{n} &= 
\int_{-1}^{1} \rmd x f(ax + b) W(x) \phi_{n}(x), \\
\Vec{e}_{n}(i) &=\phi_{n}(\hat{\mathcal{K}})  \Vec{e}(i),
\quad
\Vec{h}_{n}(i) =\phi_{n}(\hat{\mathcal{K}})  \Vec{h}(i).
\end{align}
Here, it should be noted that ${\cal T}_{n}$ {\it does not} depend on the index $i$. 
Therefore, the calculation of ${\cal T}_{n}$ can be done before any self-consistent
calculations. 
The essential mathematical relations are Eqs.~(\ref{eq:delta})-(\ref{eq:rec}). Many useful formulae about
orthogonal polynomials applicable to physical problems are found, for example, in Refs.~\cite{Nori,Lin}.

Hereafter, as shown by by Covaci {\it et al.}\cite{Covaci}, we use the Chebyshev polynomials\cite{Abramowitz;Stegun:1972}, i.e.,
\begin{subequations}
\begin{align}
\phi_{n}(x) &= \cos[n \arccos(x)],  \\
  W(x) &= \frac{1}{\sqrt{1-x^{2}}}, \quad
   w_{n}=\frac{\pi}{2}(1 + \delta_{n0}), \quad x\in [-1,1].
\end{align}
The coefficients in the recursive formula (\ref{eq:rec}) are 
\(
a_{n}=0
\), 
\(
b_{n}=2
\), and 
\(
c_{n}=1
\). 
The vector form of the formula associated with Eq.~(\ref{eq:rec_matrix}) is given as
\begin{align}
\Vec{q}_{n+1} 
= 2\hat{\mathcal{K}}  \Vec{q}_{n} -
 \Vec{q}_{n-1}
\quad
(n\ge 2), \label{eq:reccheb}
\end{align}
with $\Vec{q}_{0} = \Vec{q}$ and 
$\Vec{q}_{1} = \hat{\mathcal{K}} \Vec{q}$. 
\end{subequations} 
At the zero temperature ($\beta \to \infty$), we find that 
\begin{subequations} 
\begin{align}
& 
{\cal T}_{0} = \pi - \arccos(-b/a), \\
& 
{\cal T}_{n \neq 0} = -\frac{\sin[n \arccos(-b/a)]}{n}. \label{eq:tn}
\end{align}
\end{subequations} 
Equation (\ref{eq:reccheb}) allows the evaluation of the right hand side of Eq.~(\ref{eq:blform_d}) 
without any direct diagonalization of $\hat{\cH}$.
%
Specifically, we find that the calculations of the mean-fields (\ref{eq:cdct}) and (\ref{eq:cct}) do not require
any heavy computation in contrast to matrix diagonalization.

Finally, we remark that 
the approach shown in this section is applicable to solve 
the other equations in condensed
matter physics such as the Kohn-Sham equation in
the density functional theory with real-space
formalism.


\section{Results: Numerical Demonstrations }

We demonstrate three examples of self-consistent calculations in
inhomogeneous superconducting systems by using the above spectral-density 
polynomial expansion. 
In this section, we focus on a sigle-band $s$-wave
superconductor. 
Therefore, the quantum index $i(j)$ introduced 
in the previous section indicates a single spatial site. 
Then, the spin index $\uparrow(\downarrow)$ is separately written. 
The label ($i$, $j$) simply means a spatial 
site in real 2D space.  
Together with the BdG equations, the gap equation for $s$-wave superconductivity is
given as
\(
\Delta_{ij} = V_{ij} \langle c_{i,\downarrow} c_{j,\uparrow} \rangle.
\) 
In numerical calculations, its right hand side is expressed by using
Eq.~(\ref{eq:cct}) with truncation in the infinite summation. 
The maximum number of the summation is written as $n_{\rm c}$. 
For simplicity, we consider a simple square lattice tight-binding
model only with nearest neighbor hopping.
The hopping magnitude is $t$. 
In all simulations, we adopt $n_{\rm c} = 1000$. 
We find that this number is enough from convergent tendency except for the calculations
of the temperature dependence of $\Delta_{ij}$.

\subsection{Vortex lattice Solution}

The first example is a vortex solution and related issues.
The electromagnetic response of a superconductor under the magnetic
field has been examined on the basis of the BdG
equations\,\cite{Wang,Atkinson,Takigawa}. 
In type I\!I superconductors, the low-lying
vortex-core excitations can be studied by taking account of 
the presence of vortex lattice.

\begin{figure}[htbp]
(a)\!\includegraphics[width = 7cm]{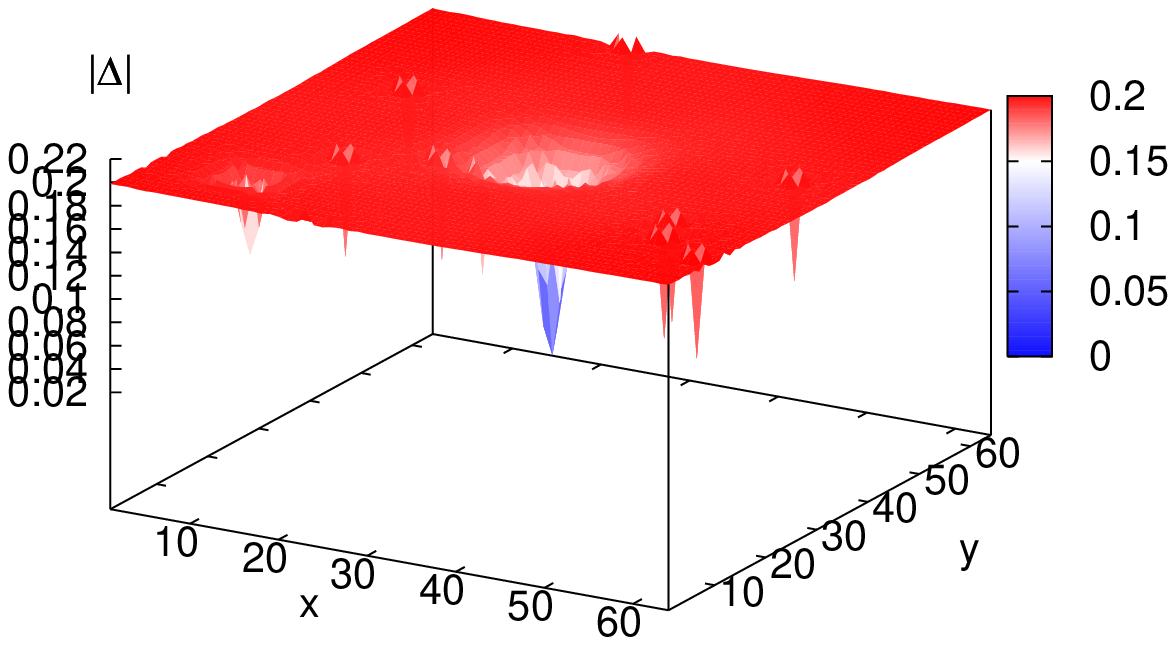}
(b)\includegraphics[width = 7cm]{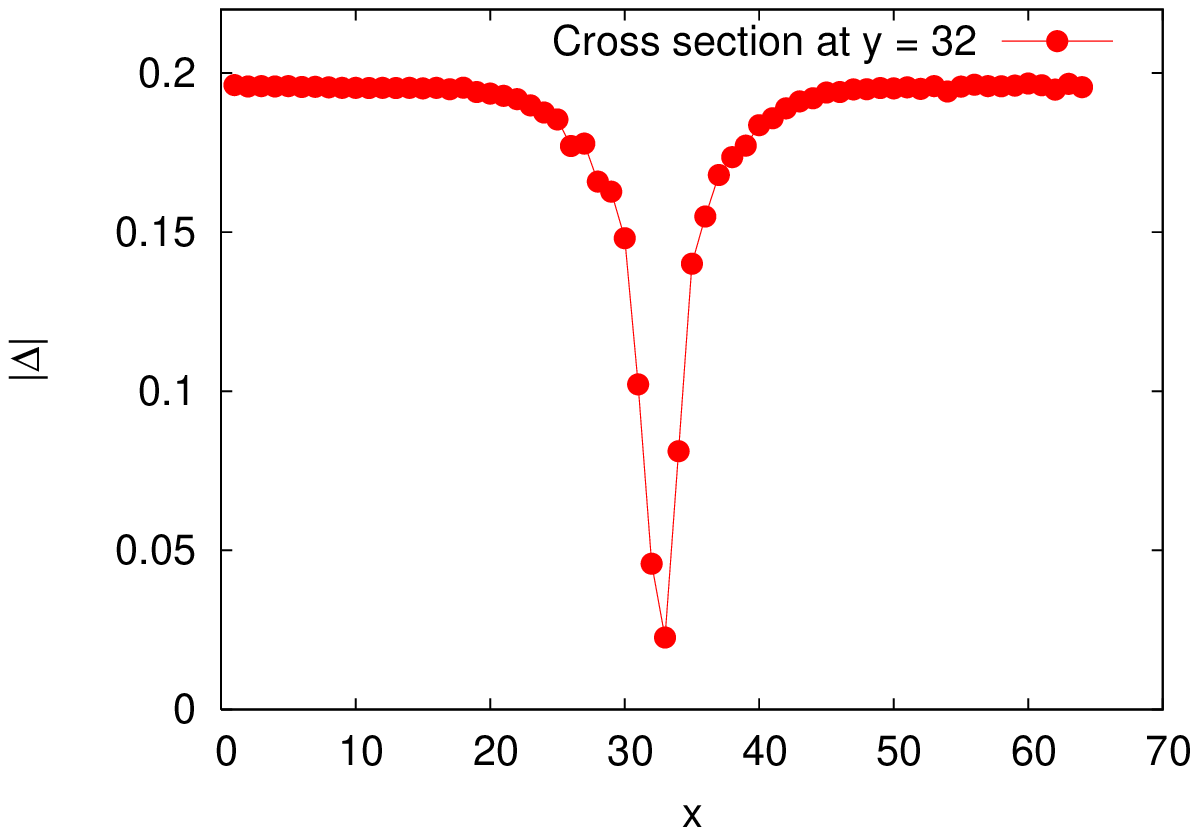}
\caption{\label{fig:vortex}
(Color online) (a) Spatial modulation of the gap amplitude in a two-dimensional vortex lattice system with ten impurities
 (b) Cross section of the spatial modulation at $y=32$. }
\end{figure}

Here, we investigate a two-dimensional $N_{x} \times N_{y}$ site system with periodic boundary condition. 
This system has a vortex square lattice.
Moreover, we introduce randomly distributed impurities 
at the zero temperature.  
The parameters are set as follows: $V_{ij} = -2.2 t \delta_{ij}$, the chemical potential 
$\mu = -1.5t$, the system size $N_{x} \times N_{y} = 64 \times 64$, and the
impurity potential $V_{\rm imp} = t$. 
We successfully calculate the gap amplitude as shown in
Fig.~\ref{fig:vortex}. 
We note that a calculation with $20$ times iterations 
takes about 40 {\it minutes} by a desktop computer 
with 8 CPU cores (Intel Xeon X5550 2.67 GHz $\times$ 2). 
Since the Hamiltonian is a sparse matrix, the calculation needs little 
computational memories. 
The separate calculations of $\Delta_{ij}$ is performed on each CPU core. 
The communication in this case, which is a one-to-all communication, is 
needed only when updating $\Delta_{ij}$. 
We confirm that these advantage points are still true in more general cases such as 
$d$-wave and multi-orbital superconductors\cite{note}.

\subsection{SNS Junction System}

The next example is the proximity effects in an SNS Josephson junction, in
which the gap symmetry of the superconducting electrodes is assumed to be $s$-wave. 
Josephson junctions and superconducting weak links have
attracted great attention in various research fields such as
superconducting device engineering \cite{Hinken:1991} including Josephson
qubits \cite{Clarke;Wilhelm:2008}, superconducting transport studies in
superconducting wires\cite{Hilgenkamp;Mannhart:2002}, fundamental studies on
unconventional nano-superconductors \cite{Hu:1994,Yamashiro;Kashiwaya:1998,Linder;Sudbo:2010,Zha;Peeters:2010}, 
and so on.  
Numerical simulations for the BdG equations has been often carried out 
to study Josephson effects\cite{MartinRodero;Yetati:1994,Yeyati;GarciaVidal:1995,Asano:2001}.  

\begin{figure}[tbp]
(a)\!\includegraphics[width = 7cm]{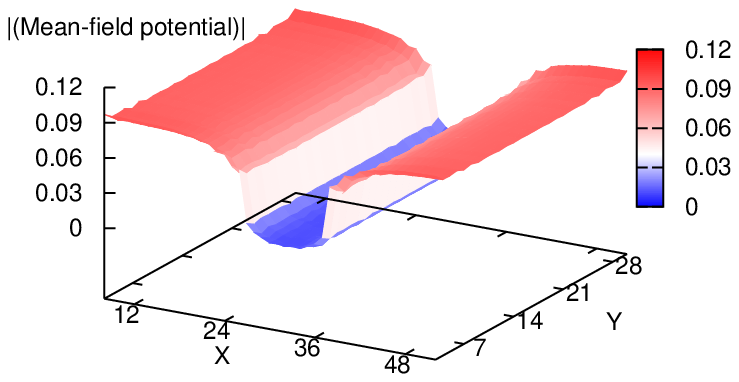}
(b)\includegraphics[width = 7cm]{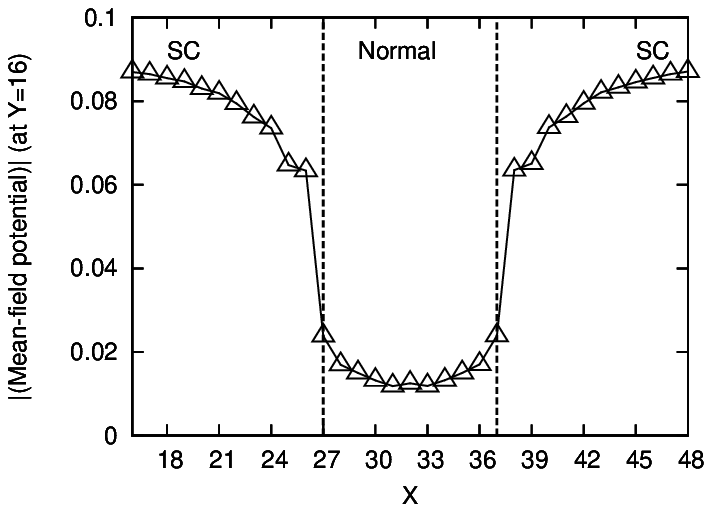}
\caption{\label{fig:sns}
(Color online) (a) Spatial modulation of mean-field potential in a
 two-dimensional SNS junction. (b) Superconducting proximity effect in the vicinity of the
 normal region at $Y=16$. }
\end{figure}

Now, let us show a spatial modulation of the mean-field, i.e., the superconducting gap 
in the SNS junction with spatial size $N_{x}\times N_{y}=64\times 32$. 
The normal conducting part is inside the region   
\(
\{(x,y); 27\le x\le 37,\,1\le y\le N_{y}\}
\). 
We set $V_{ij}=0$ inside this region. 
Otherwise, $V_{ij}=-2.2t \delta_{ij}$. 
The chemical potential $\mu=-1.5t$ and the temperature $T=0$. 
The hopping between the superconducting and normal regions
$t^{\prime}=0.8t$. 
We note that the averaged gap amplitude 
outside the normal region (i.e., in a
uniform superconductor) is $0.198$ in the present parameter set. 
It indicates that the coherence length is around $5$ sites
(Ref.\,\citen{Tinkham:1996}).  
The mean-field superconducting-gap  
\(
\langle c_{i,\uparrow}c_{i,\downarrow}\rangle
\) distribution is shown in Figs.\,\ref{fig:sns}(a) and (b). 
In this case, the self-consistent iteration number is $20$, which is confirmed to be enough through its convergence check. 
The mean-field takes a non-zero value even in the
normal region. 
We find that the spatial modulation of the mean-field is
well characterized by the length scale compatible with the coherence
length ($\sim 5$ sites).

\subsection{Temperature Dependence of Gap Amplitude}

\begin{figure}
\begin{center}
\includegraphics[width = 8cm]{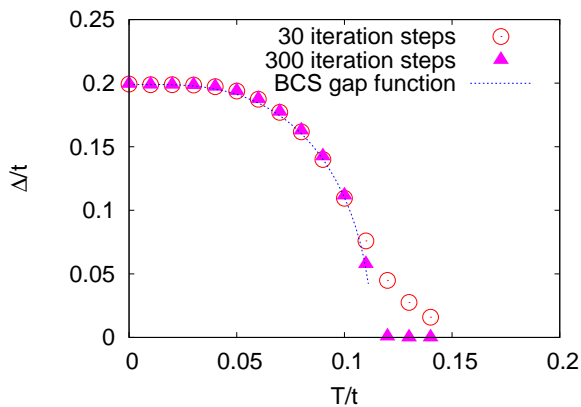}
\caption{\label{fig:temp}
(Color online) 
Temperature dependence of the gap amplitude in the two-dimensional $s$-wave superconductor.
}
\end{center}
\end{figure}

We demonstrate  temperature dependence of the superconducting gap amplitude.    
We consider a two-dimensional $N \times N$ square-plate $s$-wave superconductor.
We set $V_{ij} = -2.2 t \delta_{ij}$, the chemical potential $\mu =
-1.5t$, and the system size $N \times N = 28 \times 28$. 
The 30 and 300 times iterations are adopted to evaluate 
the temperature dependent superconducting gap, respectively. 
As shown in Fig.~\ref{fig:temp}, 
we successfully obtain the temperature dependence of the gap amplitude\cite{note2}, although 
the convergence tendency depends on the temperature range. 
These obtained temperature dependences are fitted by 
an analytical formula of the BCS gap function 
$\Delta(T) = \Delta(0) \tanh (1.74 \sqrt{\Delta(0)/(1.764 T) - 1})$. 
The slow convergence appears around $T_{\rm c}$
, since the superconducting gap almost vanishes near the critical point.
However, this is not a fault of the expansion scheme. It is known that 
the diagonalization scheme also shows similar tendency.  
A calculation with $300$ times 
iterations takes about 3 {\it minutes} when using a parallel cluster with 112 cores.

\section{Technical remarks}

In this section, we refer to technical remarks to 
describe the advantages of the present polynomial
expansion scheme. 
These points are useful when actually performing large-scale numerical calculations. 

\subsection{Matrix-Vector Product}
The calculations of $\Vec{p}^{\rm T} \Vec{q}_{n}$in Eq.~(\ref{eq:reccheb})  is necessary to evaluate mean-fields or density of states.
Since the target Hamiltonian is generally a sparse matrix, we can choose 
a fast algorithm optimized for the sparse-matrix vector product among 
several suggested ones. 
We confirm that Compressed Row Storage (CRS) format, which
is one of the typical storing-schemes 
for sparse matrices\cite{CRS}, is quite useful for the 
present calculations. 
The CRS format puts the subsequent nonzeros of the matrix row in contiguous memory locations. 
The algorithm of the matrix multiply operation $\Vec{y} = \hat{A} \Vec{x}$ is easy for the program-coding.
This algorithm is efficient on scalar processors since it has unit stride access. 
For details of the algorithm, see Ref.~\citen{CRS}. 

\subsection{Energy cut-off}
\label{subsec:choise_emin_emax}

In the present polynomial expansion scheme, 
the values of $E_{\rm max}$ and $E_{\rm min}$ are initially demanded.
A frequently-used way to do so is Lanczos\cite{Thijssen:1999}  and its relative algorithm, as pointed out in the previous paper\cite{Covaci}. 
However, we mention that we do not need any exact calculations.
Namely, what we need is just their approximate values.  
Tanaka {\it et al.} pointed out that calculation results are 
{\it insensitive} to the choice of these parameters.
%
Therefore, we simply employ the band width as the upper and lower bounds of 
the energy range as $E_{\rm max} = 10t - \mu$ and $E_{\rm min} = -10t + \mu$, or 
$E_{\rm max} = 20t - \mu$ and $E_{\rm min} = -20t + \mu$. 
In the previous section, we confirmed that this choice is effective. 

\subsection{Remarks in Mean-field Calculations}

We discuss convergence tendency of the mean-fields. 
Equations~(\ref{eq:cdct}) and
(\ref{eq:cct}) as a function of $n$ are characterized by
$\Vec{p}^{\rm T}\Vec{q}_{n}$ and ${\cal T}_{n}$, since $w_{n}$ is a
constant for the Chebyshev polynomial for $n \ge 1$.  
With the use of the unitary matrix $\hat{U}$ diagonalizing the
Hamiltonian $\hat{\cal K}$, we obtain a recurrence formula with respect to
$\Vec{x}_{n} \equiv \hat{U}^{\dagger} \Vec{q}_{n}$ as
\begin{align}
\left[ \Vec{x}_{n+1} \right]_{\gamma}&= 2  \xi_{\gamma}  \left[ \Vec{x}_{n} \right]_{\gamma} -
\left[ \Vec{x}_{n-1} \right]_{\gamma}. 
\end{align}
It indicates that 
\begin{align}
\Vec{p}^{\rm T}\Vec{q}_{n} 
&= \sum_{\mu=1}^{2 N} \sum_{\gamma=1}^{2 N} p_{\mu} U_{\mu \gamma} q_{\gamma} 
\cos \left[ n \arccos (\xi_{\gamma} ) \right].
\end{align}
Therefore, $\Vec{p}^{\rm T}\Vec{q}_{n}$ is an oscillating function with
respect to $n$.  
This is why the conversion of the local density of state (\ref{eq:ldos})
is slow, as Covaci {\it et al.} reported\cite{Covaci}.  
However, 
the conversion of the mean-fields is faster
than that of the local density of states.  
As shown in Eq.~(\ref{eq:tn}) at zero temperature, 
the $n$-dependence of ${\cal T}_{n}$ is written as 
\begin{align}
{\cal T}_{n} \propto \frac{1}{n}.
\end{align}
Then, $\Vec{p}^{\rm T}\Vec{q}_{n} {\cal T}_{n}$ shows a damped oscillating behavior on $n$. 
In the present calculations, we set $n_{\rm c}=1000$, which is rather smaller
than the value used in Ref.~\citen{Covaci}. 
On the one hand, 
we confirm that the value is enough for calculating the mean-fields. 
On the other hand, we find that the calculation of the local density of states
requires larger $n_{\rm c}$ than that of the mean-fields. 
When one tries to use the
Chebyshev polynomial expansion scheme, one should 
choose a proper value of $n_{\rm c}$ depending on 
the calculation target.
However, 
the first choice is $n_{c} = 1000$. 
Then, its tuning (e.g., $n_{c} = 500$, $2000$, $4000$, etc.) should be done.
One can easily try the various values since the computational time is proportional to the cutoff $n_{c}$.
Typically, the validity of the adopted value of $n_c$ is checked by investigating whether the resultant DOS has an oscillation behavior (i.e., Gibbs oscillation).

\subsection{ Origin of Calculation Efficiency }

\begin{figure}
\begin{center}
\includegraphics[width = 8cm]{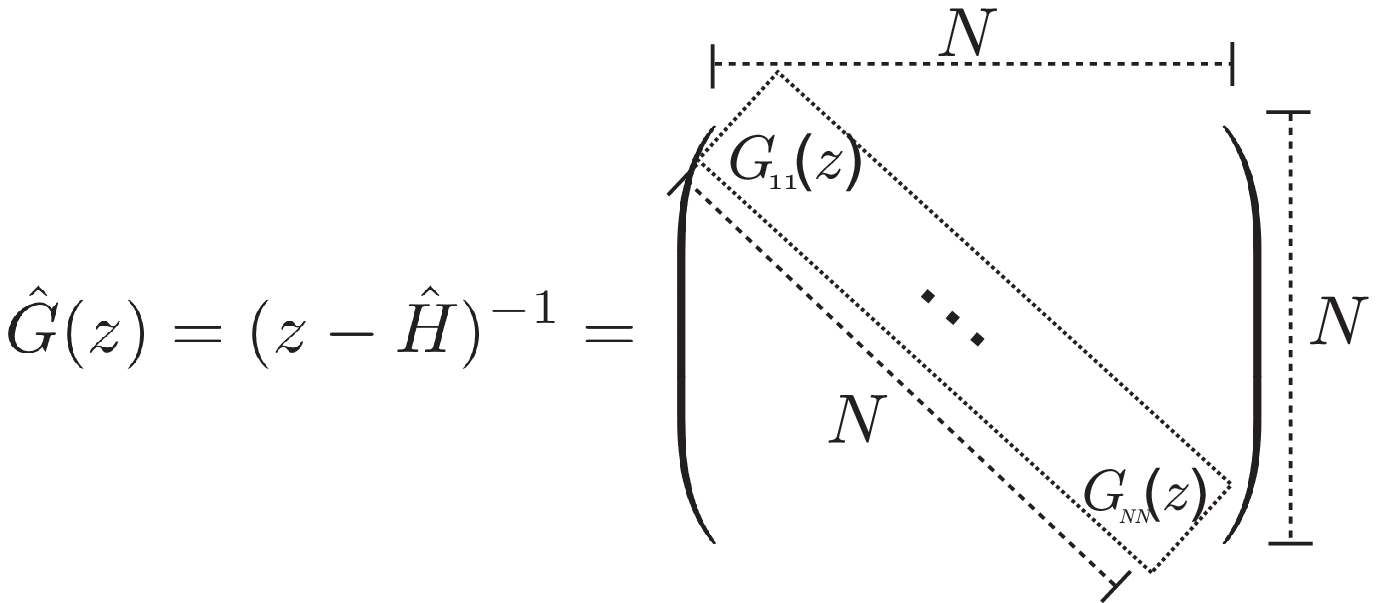}
\hspace{40mm}
\caption{\label{fig:green}
Schematic figure of $N$ elements of the $N \times N$ Green's function. 
}
\end{center}
\end{figure}

We mention a reason why the polynomial expansion scheme is much more efficient than
full-diagonalization.  
Principally, the $N \times N$ matrix Green's function is 
constructed by all eigenvectors of $\hat{\cH}$. 
The full diagonalization directly calculates them. 
On the other hand, practically, all what we have to obtain to 
solve the BdG equations with the mean-fields 
is $N$ diagonal elements of the $N \times N$ Green's function as shown in Fig.~\ref{fig:green}. 
This indicates that the calculation cost 
can be considerably reduced compared to the full diagonalization.
This is an origin of excellent efficiency of the polynomial expansion scheme. 

\begin{figure}
\begin{center}
\includegraphics[width = 7cm]{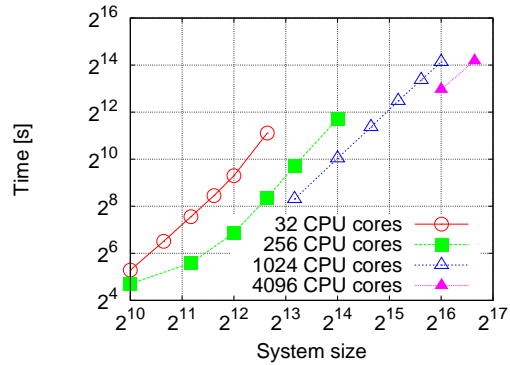}
\hspace{40mm}
\caption{\label{fig:cpus}
(Color online) 
System-size dependence of the self-consistent calculation in the $M \times M$ square lattice $s$-wave 
superconductor at zero temperature. 
The number of the iteration steps is 20. 
The system size denotes $M \times M$. 
}
\end{center}
\end{figure}

Here, let us evaluate computational costs in the present self-consistent calculations. 
We measure the elapsed time from making the Hamiltonian matrix with CSR format to 
finishing the $20$ times self-consistent iterations in the $M \times M$ square lattice 
$s$-wave superconductor at zero temperature. For the measurement, we use 
a supercomputing system PRIMERGY BX900 in Japan Atomic Energy Agency. 
As shown in Fig.~\ref{fig:cpus}, the elapsed time of 
the self-consistent calculation grows in ${\cal O}(N^{2})$ manner 
with increasing the system size $N (= M \times M)$.
The tendency is kept from 32 to 4096 cores.
In fact, the computational cost is theoretically estimated to be ${\cal O}(N^{2})$ according 
to Eq.~(21). 
Here, we mention that, although the sparse-matrix multiply operation is ${\cal O}(N)$, the mean field 
calculation on all sites requires an extra cost represented as $ {\cal O}(N) \times N$.
The previous report\cite{Covaci} claimed that the cost is ${\cal O}(N)$.
However, it is practically ${\cal O}(N^{2})$ when including 
the mean field calculation. 
In contrast, the full diagonalization scheme inevitably demands ${\cal O}(N^{3})$ cost in 
the core part of the calculation. 
This is a big advantage of the polynomial expansion
scheme. Furthermore, we focus on the speed of a self-consistent calculation by the 
present scheme.
For example, a full calculation on $256 \times 256$ ($2^{16}$) lattice system, whose matrix 
dimension size is 131077, takes about 5 hours ($\sim 2^{14}$ sec) for 20 iterations
when using 1024 CPU cores.
If we execute 20 times diagonalizations, then the elapsed time exceeds 
much over 5 hours on the same number of cores. 
The difference in CPU time becomes much more remarkable as the 
system size grows into nano to meso-scales.

\section{Conclusion}

In conclusion, we presented a full formulation of the polynomial expansion scheme 
for generic fermionic quadratic Hamiltonians with mean fields.
The spectral density in Green's function was expanded by a set of polynomials, and 
the calculation scheme of the mean fields was explicitly described. 
The scheme was actually implemented to solve the BdG equations 
for inhomogeneous superconductors.
We demonstrated three examples of nano-scale self-consistent calculations for inhomegeneous superconductors,
whose targets are a vortex lattice under randomly distributed impurities,  proximity 
induced superconducting gap in an SNS junction, and temperature dependence of the gap amplitude 
in nano-square 2D plate superconductor.
In all these calculations, we confirmed its high numerical efficiency.  
We presented technical remarks to 
take advantage of the polynomial
expansion scheme. 
These practical remarks are useful when actually performing large-scale  numerical calculations. 
The CPU cost scales with the system size $N$ in the manner ${\cal O}(N^{2})$ 
in contrast to direct full diagonalization ${\cal O}(N^{3})$, and 
its relationship becomes much more crucial in larger systems.
We claim that our scheme widely expands the calculation range and makes it possible to study 
meso-scale superconducting phenomena.

\begin{acknowledgment}

The authors would like to acknowledge Ryo Igarashi, Noriyuki Nakai and Susumu Yamada 
for helpful discussions and comments. 

\end{acknowledgment}


\begin{thebibliography}{99}
\bibitem{Bonca}
N. M. Chtchelkatchev, T. I. Baturina, A. Glatz, V. M. Vinokur, A. Omelyanchouk, and Y. Yerin: 
 {\it Physical Properties of Nanosystems} 
 edited by J. Bonca and S. Kruchinin
  (Springer, 2011) p.87-118.
\bibitem{Inhomo}
E. {\v S}im\'anek, 
{\it Inhomogeneous Superconductors: Granular and Quantum Effects}
(Oxford Univ Pr on Demand, 1994)
\bibitem{Tinkham:1996}
M. Tinkham, 
{\it Introduction to Superconductivity} 2nd ed. 
(Dover, New York, 1996)
\bibitem{Ota}
Y. Ota, M. Machida, T. Koyama, and H. Matsumoto: 
Phys. Rev. B {\bf 81} (2010) 014502.
\bibitem{deGennes:1999}
P. G. de Gennes, 
{\it Superconductivity of Metals and Alloys} 
(Westview Press, Perseus Book Group, Colorado, 1999). 
\bibitem{Covaci}
L. Covaci F. M. Peeters, and M. Berciu:   
Phys. Rev. Lett. {\bf 105} (2010) 167006.
\bibitem{Zha;Peeters:2010}
G. Q. Zha, L. Covaci, S. P. Zhou, and F. M. Peeters:  
Phys. Rev. B {\bf 82} (2010) 140502(R).
\bibitem{Han}
Q. Han, T. Li, and Z. D. Wang: Phys. Rev. B {\bf 82} (2010) 052503.
\bibitem{Weisse}
A. Wei\ss e, G. Wellein, A. Alvermann, and H. Fehske: Rev. Mod. Phys. {\bf 78} (2006) 275. 
\bibitem{Kunishima}
W. Kunishima, M. Itoh, and H. Tanaka: Prog. Theo. Phys. Supplement {\bf 138} (2000) 149. 
\bibitem{Tanaka}
H. Tanaka, W. Kunishima, and M. Itoh: RIKEN Review {\bf 29} (2000) 20. 
\bibitem{Sota}
S. Sota and M. Itoh: J. Phys. Soc. Jpn. {\bf 76} (2007) 054004. 
\bibitem{Okumura}
S. Zhang, S. Yamagiwa, M. Okumura, and S. Yunoki, 
in  {\it IPDPS/APDCM 2011}, 2011, (Anchorage USA, 2011), p. 564-571.
\bibitem{NagaiQPI}
Y. Nagai, N. Nakai, and M. Machida, arXiv:1103.5842. 
\bibitem{vanHemmen:1980}
J. L. van Hemmen: Z. Phys. B {\bf 38} (1980) 271.
\bibitem{Mahan:2000}
G. D. Mahan, 
{\it Many-particle Physics} 3rd ed. 
(Kluwer Academic/Plenum Publishers, New York, 2000).
\bibitem{Abramowitz;Stegun:1972}
{\it Handbook of Mathematical Functions with Formulas, Graphs, and
	Mathematical Tables}, 
edited by M. Abramowitz and I. A. Stegun 
(Dover, New York, 1972), Chap.22.
\bibitem{Nori}
F. Nori and Y.-L. Lin: 
Phys. Rev. B {\bf 49} (1994) 4131.
\bibitem{Lin}
Y.-L. Lin and F. Nori: 
Phys. Rev. B {\bf 50} (1994) 15953.
\bibitem{Wang}
Y. Wang and A. H. MacDonald: 
Solid State Commun. {\bf 109} (1998) 289.
\bibitem{Atkinson}
W. A. Atkinson and A. H. MacDonald:  
Phys. Rev. B {\bf 60} (1999) 9295.
\bibitem{Takigawa}
M. Takigawa, M. Ichioka, and K. Machida:  
J. Phys. Soc. Jpn. {\bf 12} (2000) 3943. 
\bibitem{note}
The impurity problem in a $d$-wave superconductors are studied in preparation.
\bibitem{Hinken:1991}
J. H. Hinken, 
{\it Superconductor Electronics: Foundations and Microwave Applications} 
(Springer-Verlag, Berlin, 1991).
\bibitem{Clarke;Wilhelm:2008}
J. Clarke and F. K. Wilhelm:  
Nature {\bf 453} (2008) 1031.
\bibitem{Hilgenkamp;Mannhart:2002}
H. Hilgenkamp and J. Mannhart:  
Rev. Mod. Phys. {\bf 74} (2002) 485. 
\bibitem{Hu:1994}
C. R. Hu: 
Phys. Rev. Lett. {\bf 72} (1994) 1526. 
\bibitem{Yamashiro;Kashiwaya:1998}
M. Yamashiro, Y. Tanaka, and S. Kashiwaya:  
J. Phys. Soc. Jpn. {\bf 67} (1998) 3364.
\bibitem{Linder;Sudbo:2010}
J. Linder, A. M. Black-Schaffer, A. Sudb\o: 
Phys. Rev. B {\bf 82} (2010) 041409. 
\bibitem{MartinRodero;Yetati:1994}
A. Martin-Rodero, F. J. Garcia-Vidal, and A. L. Yeyati:  
Phys. Rev. Lett. {\bf 72} (1994) 554. 
\bibitem{Yeyati;GarciaVidal:1995}
A. L. Yeyati, A. Martin-Rodero, and F. J. Garcia-Vidal:  
Phys. Rev. B {\bf 51} (1995) 3743. 
\bibitem{Asano:2001}
Y. Asano:  
Phys. Rev. B {\bf 63} (2001) 052512. 
\bibitem{note2}
The gap amplitude near $T_{c}$ is differ from the BCS gap function since
	the number of the self-consistent iterations steps we adopt is
	not enough around $T_{c}$.  
\bibitem{CRS}
R. Barrett, M. Berry,  T. F. Chan, J. Demmel, J. Donato, 
J. Dongarra, V. Eijkhout, R. Pozo, C. Romine, and H. Van der Vorst, 
{\it Templates for the Solution of Linear Systems: Building Blocks for Iterative Methods, } 2nd Edition. 
(SIAM, Philadelphia, PA, 1994)
\bibitem{Thijssen:1999}
J. M. Thijssen, 
{\it Computational Physics} 
(Cambridge University Press, Cambridge, England, 1999). 




\end{thebibliography}
\end{document}